\documentclass[iop,apjl,12pt]{emulateapj}
\usepackage{times}
\usepackage[normalem]{ulem}
\usepackage{apjfonts}
\usepackage{epsfig}
\usepackage{natbib}
\usepackage{amsfonts}
\usepackage{amsmath}
\usepackage{multirow}
\usepackage{enumerate}
\usepackage[usenames]{color}
\usepackage[plainpages=false, colorlinks=true, anchorcolor=lblui, linkcolor=blue, citecolor=black, urlcolor=blue, bookmarks=false]{hyperref}
\def\Dwa{$\,$\uppercase\expandafter{\romannumeral5}$\,$}

\def\sless{\lower2pt\hbox{$\buildrel {\scriptstyle <}
   \over {\scriptstyle\sim}$}}

\def\sgreat{\lower2pt\hbox{$\buildrel {\scriptstyle >}
   \over {\scriptstyle\sim}$}}
\def\sharpnull#1{}

\newcommand{\shortauth}{M{\"o}sta \emph{et al.}}
\newcommand{\slugcom}{Draft version - \today}
\slugcomment{\slugcom}

\lefthead{\sc \footnotesize \slugcom \hfill \shortauth}
\righthead{\sc \footnotesize \slugcom \hfill \shortauth}

\begin{document}
\slugcomment{Draft version \today.}


\title{Magnetorotational Core-Collapse Supernovae in Three Dimensions}

\author{Philipp M{\"o}sta\altaffilmark{1}}
\author{Sherwood Richers\altaffilmark{1}}
\author{Christian D. Ott\altaffilmark{1,2,+}}
\author{Roland Haas\altaffilmark{1}}
\author{Anthony L. Piro\altaffilmark{1}}
\author{Kristen Boydstun\altaffilmark{1}}
\author{Ernazar Abdikamalov\altaffilmark{1}}
\author{Christian Reisswig\altaffilmark{1,++}}
\author{Erik Schnetter\altaffilmark{3,4,5}}
  \altaffiltext{1}{TAPIR, Mailcode 350-17,
  California Institute of Technology, Pasadena, CA 91125, USA, 
  pmoesta@tapir.caltech.edu}
\altaffiltext{2}{Kavli Institute for the Physics and
 Mathematics of the Universe (Kavli IPMU WPI), The University of Tokyo, Kashiwa, Japan}
\altaffiltext{3}{Perimeter Institute for Theoretical Physics, Waterloo, ON, Canada.}
\altaffiltext{4}{Department of Physics, University of Guelph, Guelph, ON, Canada.}
\altaffiltext{5}{Center for Computation \& Technology, Louisiana State University, Baton Rouge, USA.}
\altaffiltext{+}{Alfred P. Sloan Research Fellow}
\altaffiltext{++}{NASA Einstein Fellow}

\begin{abstract}
We present results of new three-dimensional (3D) general-relativistic
magnetohydrodynamic simulations of rapidly rotating strongly
magnetized core collapse. These simulations are the first of their
kind and include a microphysical finite-temperature equation of state
and a leakage scheme that captures the overall energetics and lepton
number exchange due to postbounce neutrino emission. Our results show
that the 3D dynamics of magnetorotational core-collapse supernovae are
fundamentally different from what was anticipated on the basis of
previous simulations in axisymmetry (2D). A strong bipolar jet that
develops in a simulation constrained to 2D is crippled by a spiral
instability and fizzles in full 3D. While multiple
(magneto-)hydrodynamic instabilities may be present, our analysis
suggests that the jet is disrupted by an $m=1$ kink instability of the
ultra-strong toroidal field near the rotation axis. Instead of an
axially symmetric jet, a completely new, previously unreported flow
structure develops. Highly magnetized spiral plasma funnels expelled
from the core push out the shock in polar regions, creating wide
secularly expanding lobes. We observe no runaway explosion 
by the end of the full 3D simulation at $185\,\mathrm{ms}$ after bounce. At
this time, the lobes have reached maximum radii of $\sim$$900\,\mathrm{km}$.

\end{abstract}

\section{Introduction}

\begin{figure*}[t]
\includegraphics[width=0.247125\textwidth]{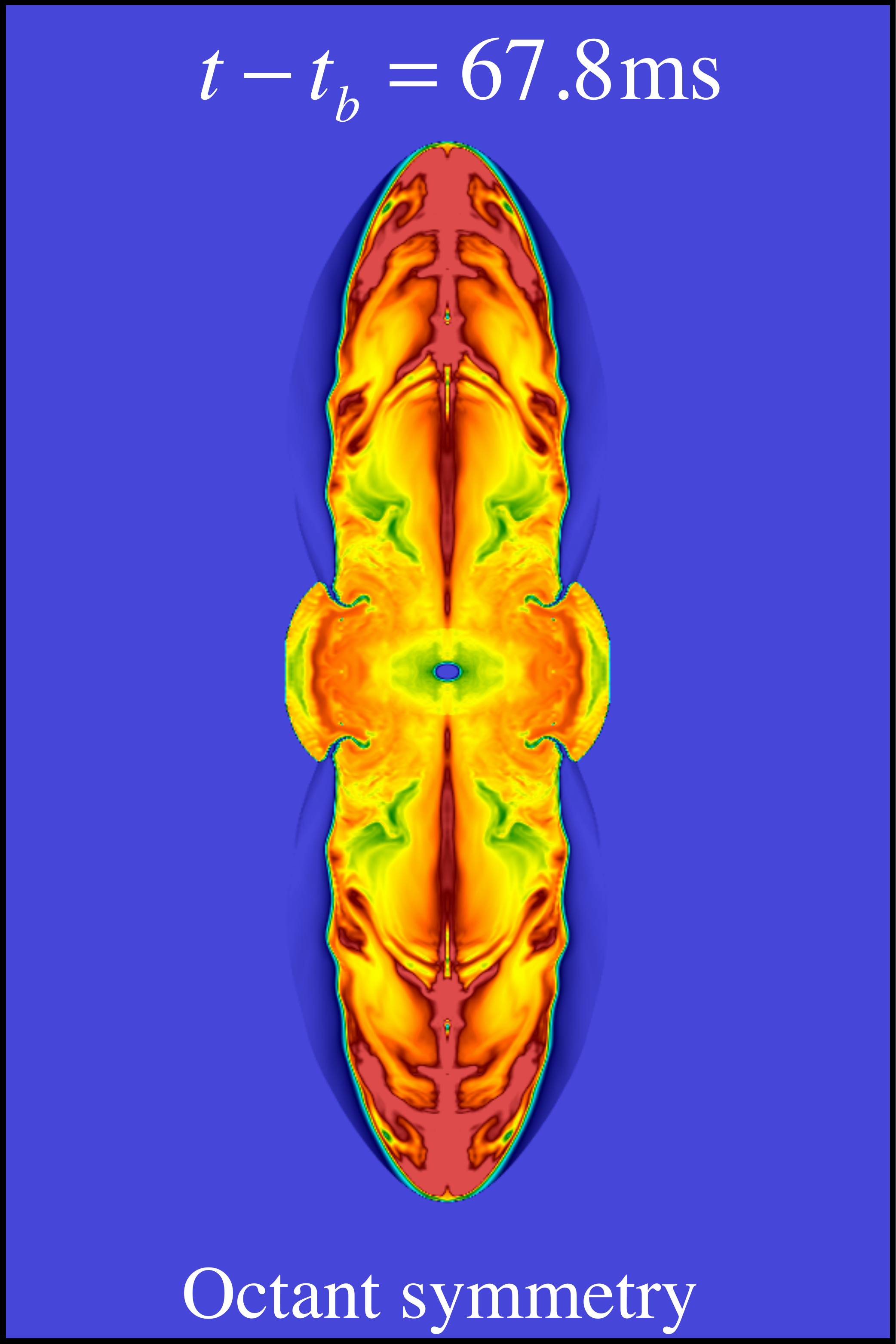}
\includegraphics[width=0.247125\textwidth]{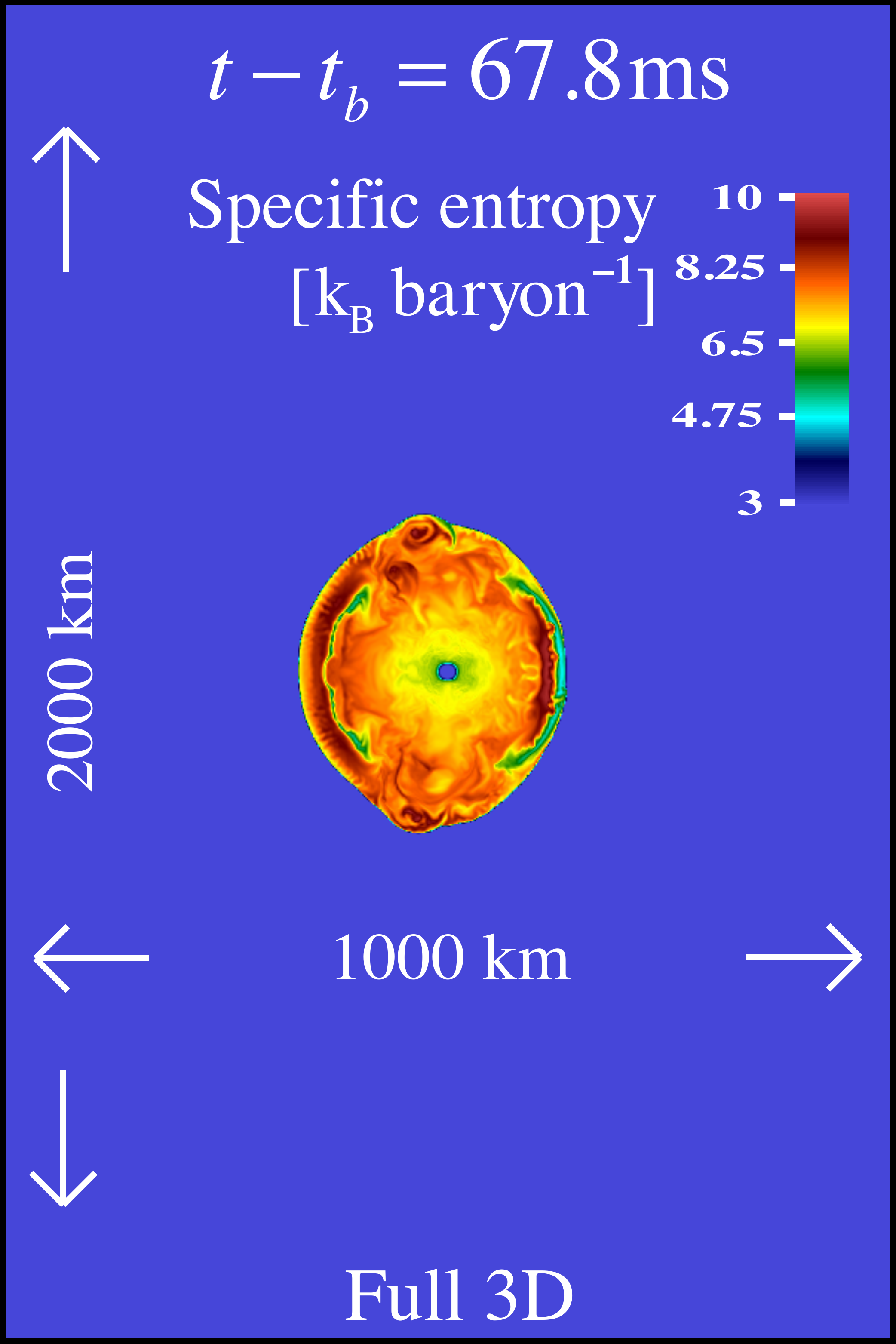}
\includegraphics[width=0.247125\textwidth]{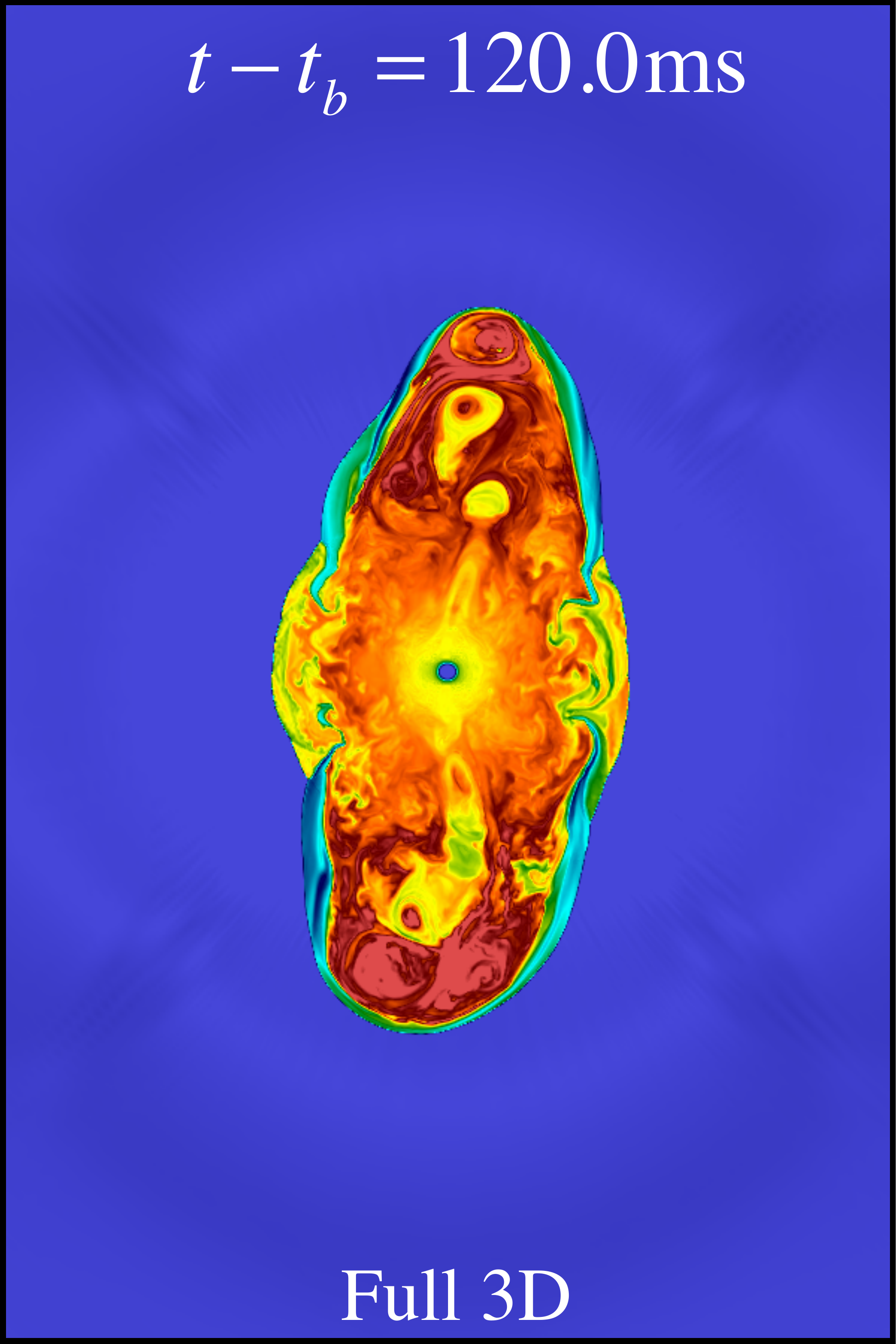}
\includegraphics[width=0.247125\textwidth]{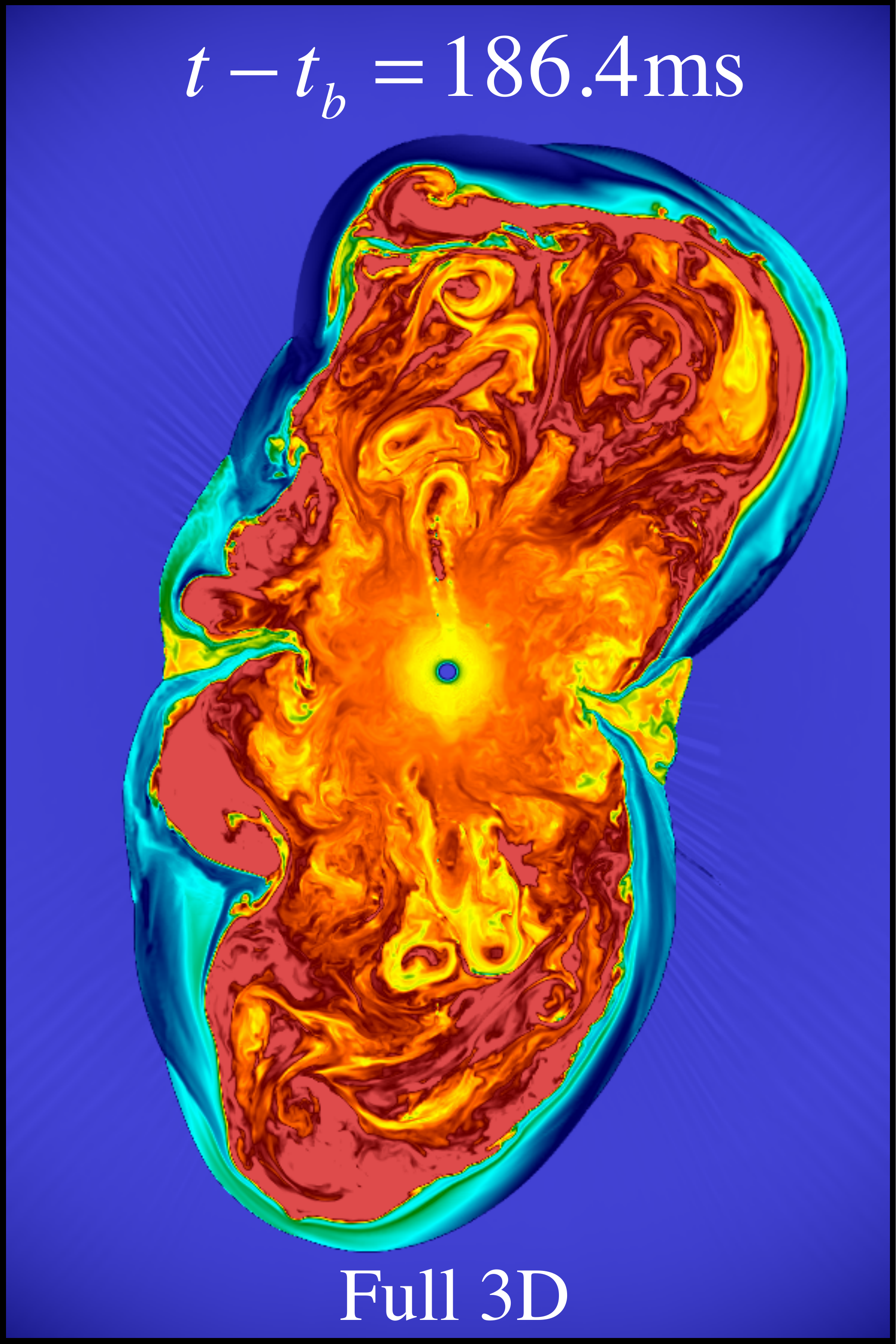}
\caption{Meridional slices ($x-z$ plane; $z$ being the vertical) of
  the specific entropy at various postbounce times. The ``2D'' (octant
  3D) simulation (leftmost panel) shows a clear bipolar jet, while in
  the full 3D simulation (3 panels to the right) the initial jet fails
  and the subsequent evolution results in large-scale asymmetric
  lobes.}
\label{fig:octfullcmp} \vspace{0.5cm} \end{figure*}

Stellar collapse liberates gravitational energy of order
$10^{53}\,\mathrm{erg\,s}^{-1}$ ($100\, \mathrm{B}$).  Most ($99\%$)
of that energy is emitted in neutrinos, and the remainder ($\lesssim
1\, \mathrm{B}$) powers a core-collapse supernova (CCSN) explosion.
However, a small fraction of CCSNe are hyper-energetic ($\sim 10\,
\mathrm{B}$) and involve relativistic outflows (e.g.,
\citealt{soderberg:06,drout:11}). These \emph{hypernovae} come from
stripped-envelope progenitors and are classified as Type Ic-bl (H/He
deficient, broad spectral lines).  Importantly, all SNe connected with
long gamma-ray bursts (GRB) are of Type Ic-bl
\citep{modjaz:11,hjorth:11}.

Typical $\mathcal{O}(1) \mathrm{B}$ SNe may be driven by the {\it
  neutrino mechanism} \citep{bethe:85}, in which neutrinos emitted
from the collapsed core deposit energy behind the stalled shock,
eventually driving it outward (e.g., \citealt{mueller:12a,bruenn:13}).
However, the neutrino mechanism appears to lack the efficiency needed
to drive hyperenergetic explosions. One possible alternative is the
\emph{magnetorotational} mechanism (e.g.
\citealt{bisno:70,leblanc:70,meier:76,wheeler:02}). In its canonical
form, rapid rotation of the collapsed core (Period
$\mathcal{O}(1)\,\mathrm{ms}$, spin energy
$\mathcal{O}(10)\,\mathrm{B}$) and magnetar-strength magnetic field
with a dominant toroidal component 
drive a strong bipolar jet-like explosion that could result in a
hypernova (\citealt{burrows:07b}).

The magnetorotational mechanism requires rapid precollapse rotation
($P_0 \lesssim 4\,\mathrm{s}$; \citealt{ott:06spin,burrows:07b}) and
an efficient process to rapidly amplify the likely weak seed magnetic
field of the progenitor. The \emph{magnetorotational instability}
(MRI,~\citealt{balbus:91,akiyama:03,obergaulinger:09}) is one
possibility.  The MRI operates on the free energy of differential
rotation and, in combination with dynamo action, has been hypothesized
to provide the necessary global field strength on an essentially
dynamical timescale \citep{akiyama:03,thompson:05}. The wavelength of
the fastest growing MRI mode in a postbounce CCSN core is much smaller
than what can currently be resolved in global multi-dimensional CCSN
simulations.  Under the assumption that MRI and dynamo operate as
envisioned, a common approach is to start with a likely unphysically
strong precollapse field of $10^{12}-10^{13}\,\mathrm{G}$.  During
collapse and the early postbounce evolution, this field is amplified
by flux compression and rotational winding to dynamically important
field strength of $B_\mathrm{tor} \gtrsim 10^{15}-10^{16}\,\mathrm{G}$
\citep{burrows:07b}. 
In this way, a number of recent two-dimensional
(2D) magnetohydrodynamic (MHD) simulations have found robust and
strong jet-driven explosions (e.g.,
\citealt{shibata:06,burrows:07b,takiwaki:11}). Only a handful of 3D
studies have been carried out with varying degrees of microphysical
realism (\citealt{mikami:08,kuroda:10,scheidegger:10b,winteler:12})
and none have compared 2D and 3D dynamics directly.


In this \emph{Letter}, we present new full 3D dynamical-spacetime
general-relativistic MHD (GRMHD) simulations of rapidly rotating
magnetized CCSNe. These are the first to employ a microphysical
finite-temperature equation of state, a realistic progenitor model,
and an approximate neutrino treatment for collapse and postbounce
evolution. We carry out simulations in full unconstrained 3D and
compare with simulations starting from identical initial
conditions, but constrained to 2D.
Our results for a model with initial poloidal $B$-field of
$10^{12}\,\mathrm{G}$ indicate that 2D and 3D magnetorotational CCSNe
are fundamentally different. In 2D, a strong jet-driven explosion
obtains, while in unconstrained 3D, the developing jet is destroyed by
nonaxisymmetric dynamics, caused most likely by an $m=1$
MHD kink instability. The subsequent CCSN evolution leads to two large
asymmetric shocked lobes at high latitudes.  Highly-magnetized tubes
tangle, twist, and drive the global shock front steadily, but not
dynamically outward. A runaway explosion does not occur during the
$\sim$$185\,\mathrm{ms}$ of postbounce time covered by our full 3D
simulation.

\section{Methods and Setup}

We employ ideal GRMHD with adaptive mesh refinement (AMR) and
spacetime evolution provided by the open-source
\texttt{EinsteinToolkit} ~\citep{moesta:14a,loeffler:12}. GRMHD is
implemented in a finite-volume fashion with WENO5
reconstruction~\citep{reisswig:12, tchekhovskoy:07} and the HLLE
Riemann solver \citep{HLLE:88} and constrained transport
\citep{toth:00} for maintaining $\mathrm{div} B = 0$. We employ the
$K_0 = 220\,\mathrm{MeV}$ variant of the equation of state of
\cite{lseos:91} and the neutrino leakage/heating approximations
described in \cite{oconnor:10} and \cite{ott:12a}.  At the precollapse
stage, we cover the inner $\sim$$5700\,\mathrm{km}$ of the star with
four AMR levels and add five more during collapse. After
bounce, the protoneutron star is covered with a resolution of
$\sim$$370\,\mathrm{m}$ and AMR is set up to always cover the shocked
region with at least $1.48\,\mathrm{km}$ linear resolution.

 
We draw the $25$-$M_\odot$ (at zero-age-main-sequence) presupernova
model E25 from \cite{heger:00} and set up axisymmetric precollapse
rotation using the rotation law of \cite{takiwaki:11} (see their
Eq.~1) with an initial central angular velocity of $2.8\,
\mathrm{rad}\, \mathrm{s}^{-1}$. The fall-off in cylindrical radius
and vertical position is controlled by parameters $x_0 =
500\,\mathrm{km}$ and $z_0 = 2000\,\mathrm{km}$, respectively. 
We set up the initial magnetic field by a vector potential of the form
$A_r = A_\theta = 0; A_\phi = B_0 ({r_0^3})({r^3+r_0^3})^{-1}\, r \sin
\theta$, where $B_0$ controls the strength of the field.
In this way we obtain a modified dipolar field structure that stays
nearly uniform in strength within radius $r_0$ and falls off like a
dipole at larger radii. We set $B_{0} = 10^{12}\, \mathrm{G}$ and
choose $r_0 = 1000\, \mathrm{km}$ to match the initial conditions of
model B12X5$\beta$0.1 of the 2D study of \cite{takiwaki:11}, in which
a jet-driven explosion is launched $\sim$$20\,\mathrm{ms}$ after
bounce.

We perform simulations both in full, unconstrained 3D and in octant
symmetry 3D (90-degree rotational symmetry in the $x-y$ plane and 
reflection symmetry across the $x-y$ plane) with otherwise identical
setups. Octant symmetry suppresses most nonaxisymmetric dynamics,
since it allows only modes with azimuthal numbers that are multiples
of $m=4$. In order to study the impact of potential low-mode
nonaxisymmetric dynamics on jet formation, we add a $1\%$ $m=1$
perturbation to the full 3D run.  Focusing on a potential instability
of the strong toroidal field near the spin axis, we apply this
perturbation to the velocity field within a cylindrical radius of
$15\,\mathrm{km}$ and outside the protoneutron star, $30\,\mathrm{km} \le |z|
\le 75\,\mathrm{km}$, at $5\,\mathrm{ms}$ after bounce.


\begin{figure}[h!] \centering
\includegraphics[width=0.95\columnwidth]{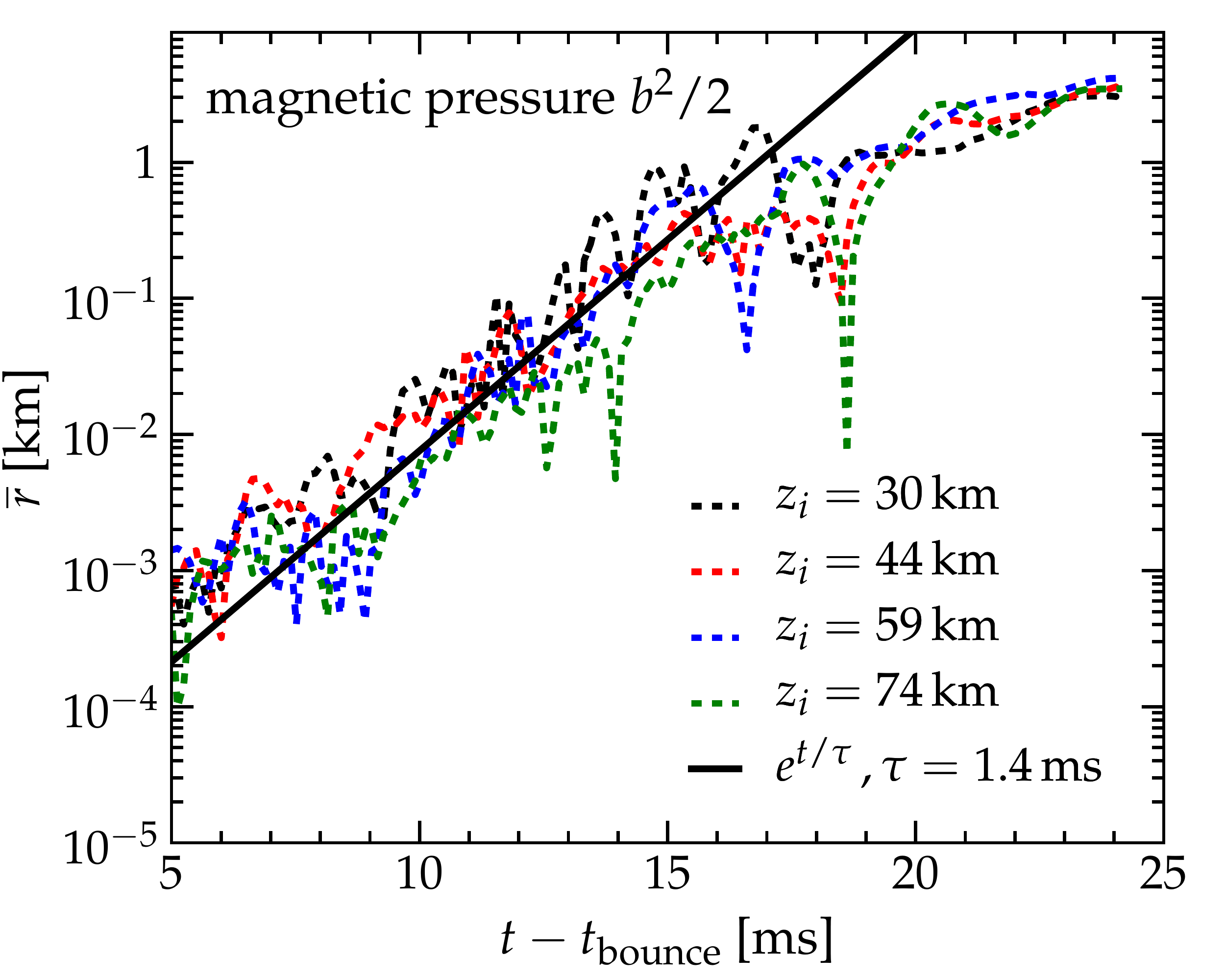}
\hspace*{-0.1cm}\includegraphics[width=0.91\columnwidth]{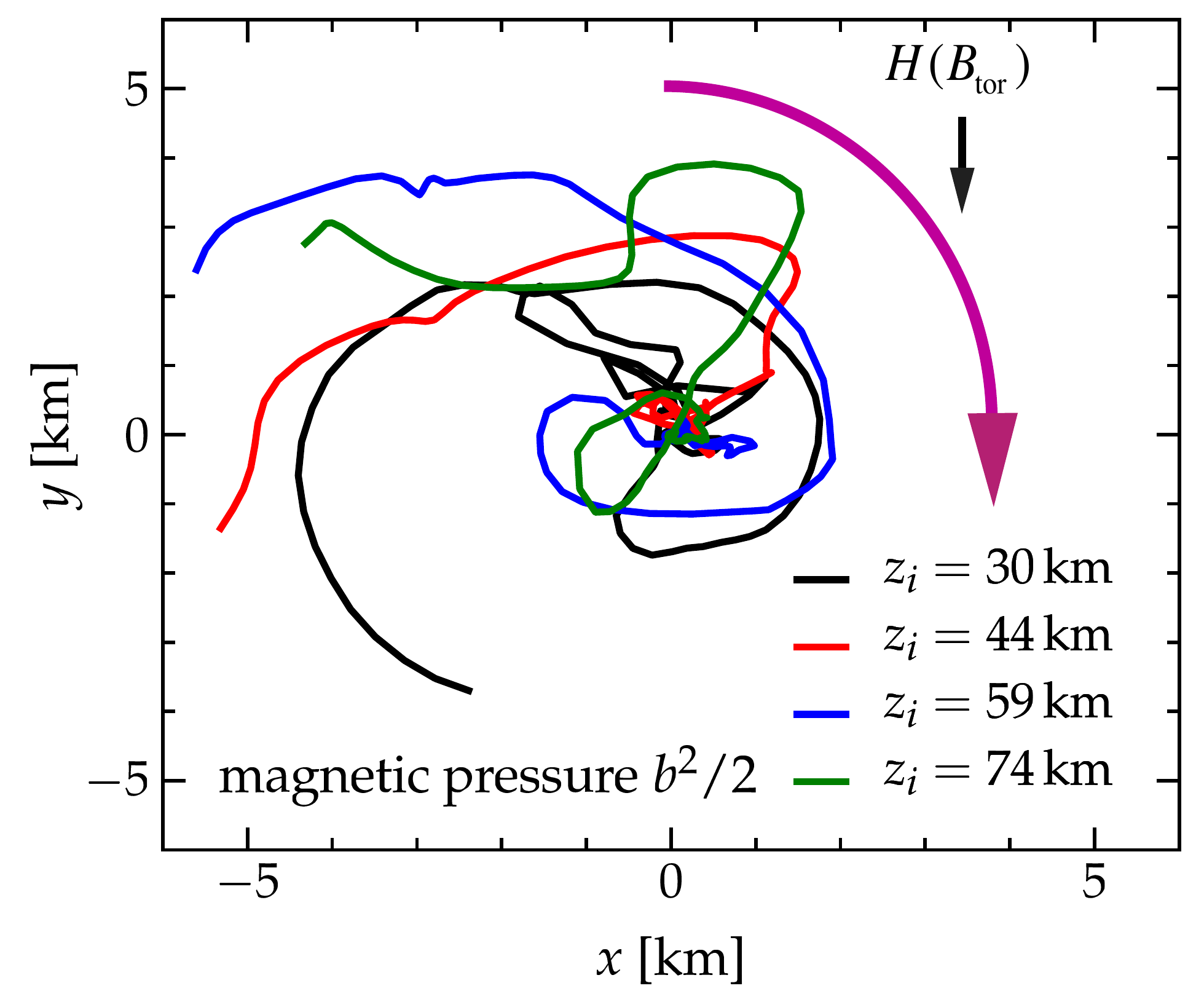} 
\caption{{\bf Top panel}: Barycenter displacement $\overline{r}$ of
  $b^2$ in $x-y$ planes at different heights $z_i$. To minimize the
  influence of material that does not belong to the unstable jet, we
  include only cells inside a radius of $15\, \mathrm{km}$. We observe
  exponential growth of the displacement in the early postbounce
  evolution until saturation at $t-t_b \sim 20\, \mathrm{ms}$.  The
  growth rate $\tau_{\textit{fgm},sim} \approx 1.4\, \mathrm{ms}$ is consistent
  with estimates for the MHD kink instability (see Section
  \ref{sec:instabilities}).  {\bf Bottom panel}: Tracks of the
  barycenter in the $x-y$ plane at different $z_i$. They are plotted
  for the interval shown in the top panel. The tracks trace the spiral
  nature of the displacement. Note that, as required for the
  perturbation to be unstable \citep{begelman:98}, the helicity of the
  displacement motion (counter-clockwise in the $x-y$ plane) is
  opposite to the helicity of the toroidal magnetic field
  $H(B_{\mathrm{tor}})$ (clockwise in the $x-y$ plane), 
  indicated by the magenta arrow.}
\label{fig:growthrate} \vspace{0.5ex} \end{figure}

\section{Results}
Collapse and the very early postbounce evolution proceed identically
in octant symmetry and full 3D. At bounce, $\sim$$350\,\mathrm{ms}$
after the onset of collapse, the poloidal and toroidal B-field
components reach $B_{\mathrm{pol}},B_{\mathrm{tor}} \sim 10^{15}\, \mathrm{G}$. The 
hydrodynamic shock launched at bounce, still approximately spherical, stalls
after $\sim$$10\,\mathrm{ms}$ at a radius of $\sim$$110\,\mathrm{km}$.
Rotational winding, operating on the extreme differential rotation
in the region between inner core and shock, amplifies the toroidal component to
$10^{16}\,\mathrm{G}$ near the rotation axis within $\sim$$20\,\mathrm{ms}$ of
bounce. At this time, the strong polar magnetic pressure gradient, in
combination with hoop stresses excerted by the toroidal field, launches a
bipolar outflow. As depicted by the leftmost panel of
Fig.~\ref{fig:octfullcmp}, a jet develops and reaches $\sim$$800\,\mathrm{km}$
after $\sim$$70\,\mathrm{ms}$ in the octant-symmetry run. The expansion speed
at that point is mildly relativistic ($v_r \simeq 0.1-0.15\, c$). This is
consistent with the 2D findings of \cite{takiwaki:11}.

The full 3D run begins to diverge from its more symmetric counterpart
around $\sim$15$\,\mathrm{ms}$ after bounce. A nonaxisymmetric spiral
($m=1$) deformation develops near the rotation axis. It distorts and
bends the initially nearly axisymmetrically developing jet, keeping it
from breaking out of the stalled shock. The nearly prompt
magnetorotational explosion of the octant-symmetry run \emph{fails} in
full 3D. The subsequent 3D evolution is fundamentally different from
2D, as evidenced by the three panels of Fig.~\ref{fig:octfullcmp}
depicting meridional specific entropy slices at different times in the
full 3D run. Until $80\,\mathrm{ms}$, the shock remains stalled and
nearly spherical. The $m=1$ dynamics pervade the entire postshock
region and cause a spiral-sloshing of the shock front that is
reminiscent of the standing-accretion shock instability in rotating 3D
CCSNe (cf.~\citealt{kuroda:13}). Later, highly-magnetized ($\beta =
P_\mathrm{gas} / P_\mathrm{mag} \ll 1$) funnels of high-entropy
material protrude from polar regions of the core and secularly push
out the shock into two dramatic tilted lobes. At the end of our
simulation ($\sim$$185\,\mathrm{ms}$ after bounce) the lobes fill
polar cones of $\sim$$90^\circ$ and are only gradually expanding as
low-$\beta$ material is pushed out from below. Accreting material is
deflected by these lobes and pushed towards the equator where it
accretes through the remainder of the initial nearly spherical shock.

\begin{figure*}[t] 
\centering
\includegraphics[width=0.1625\textwidth]{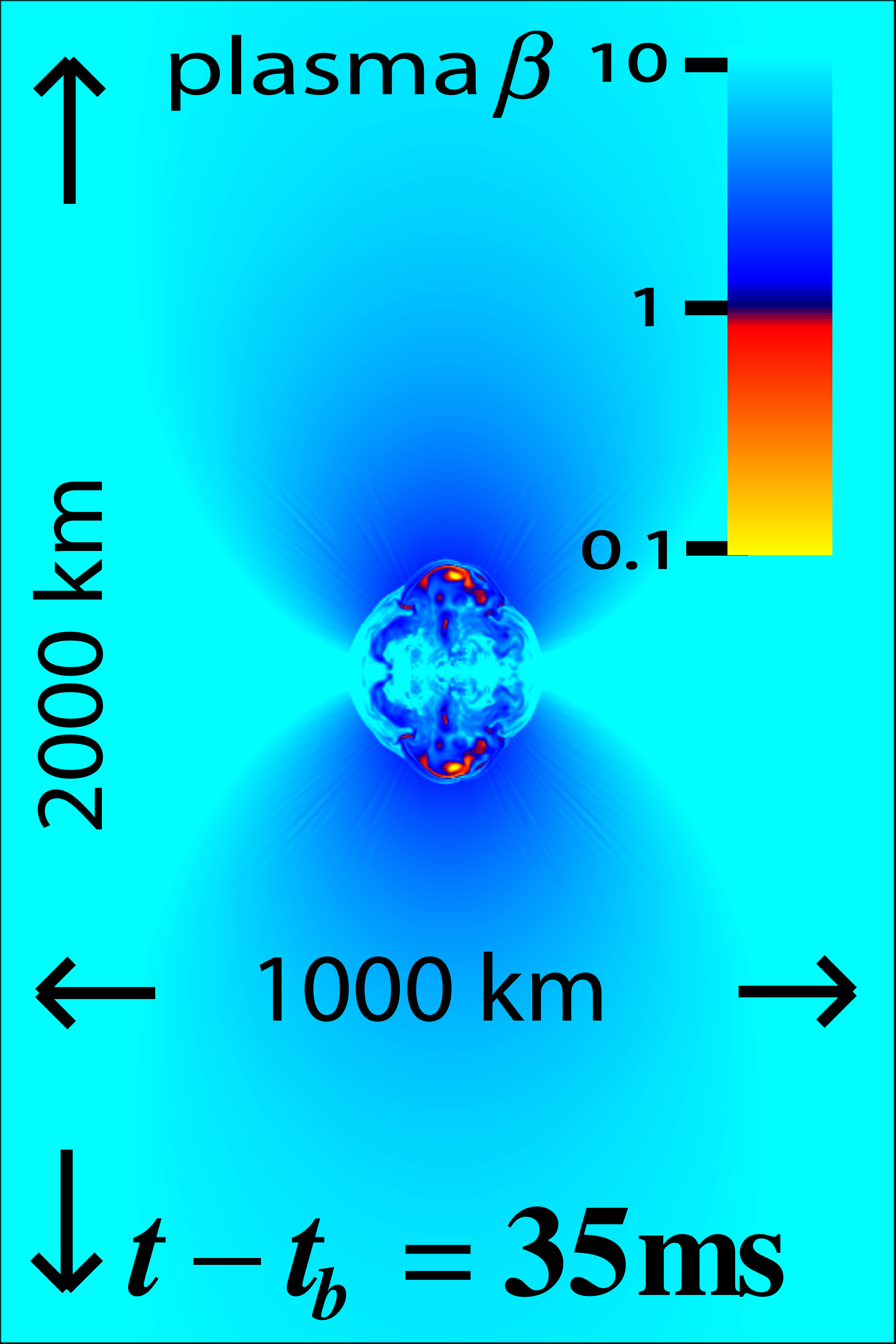}
\includegraphics[width=0.1625\textwidth]{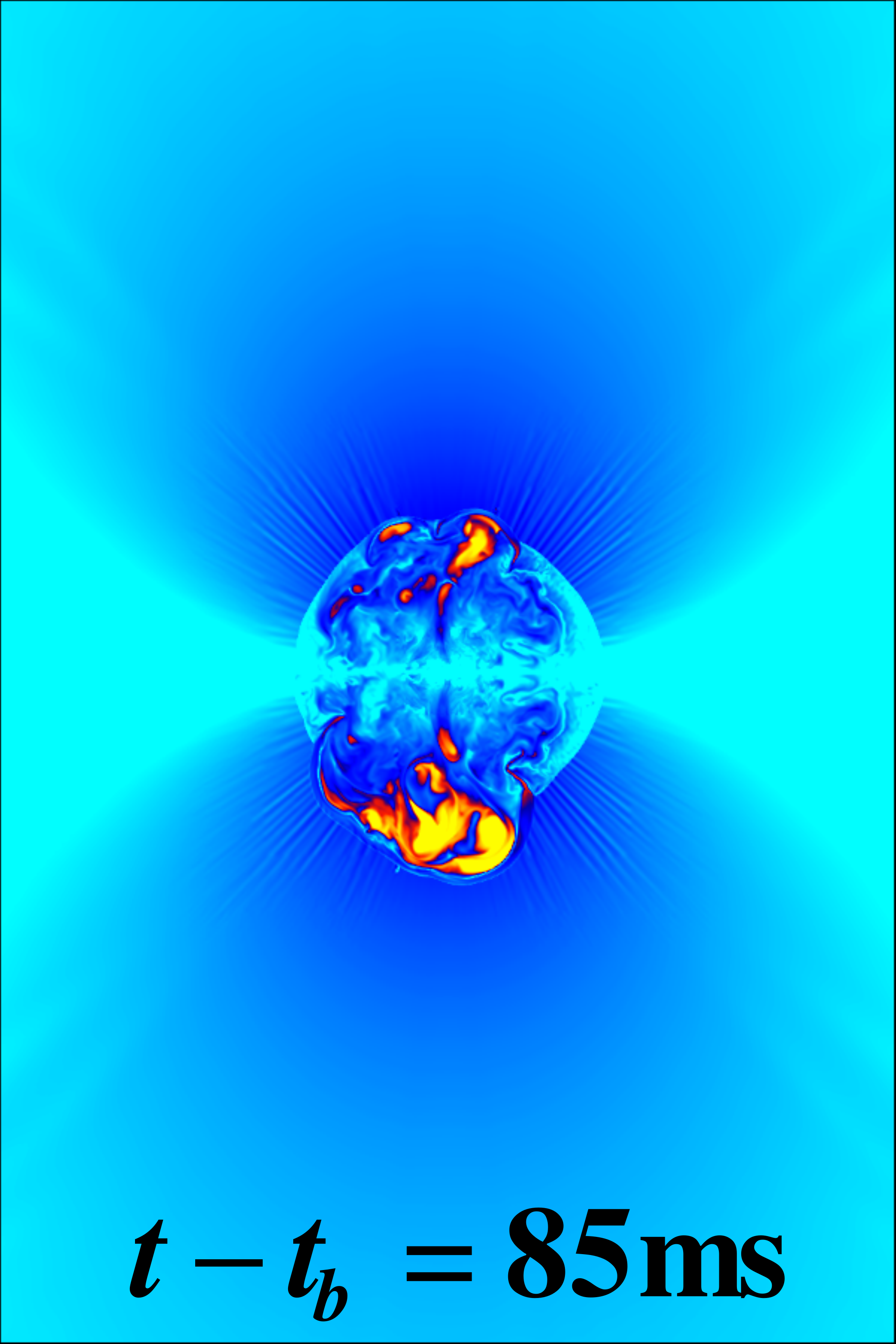}
\includegraphics[width=0.1625\textwidth]{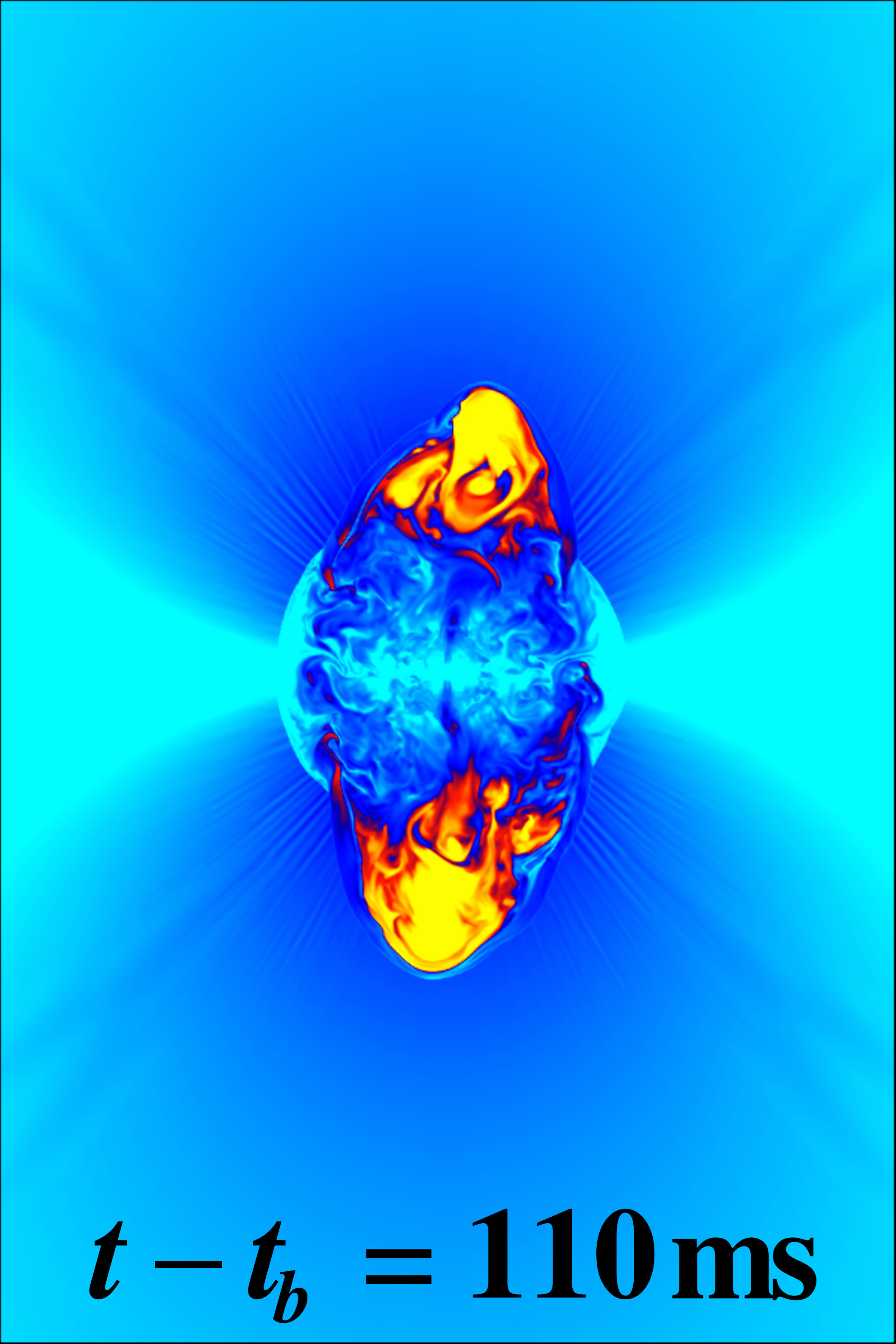}
\includegraphics[width=0.1625\textwidth]{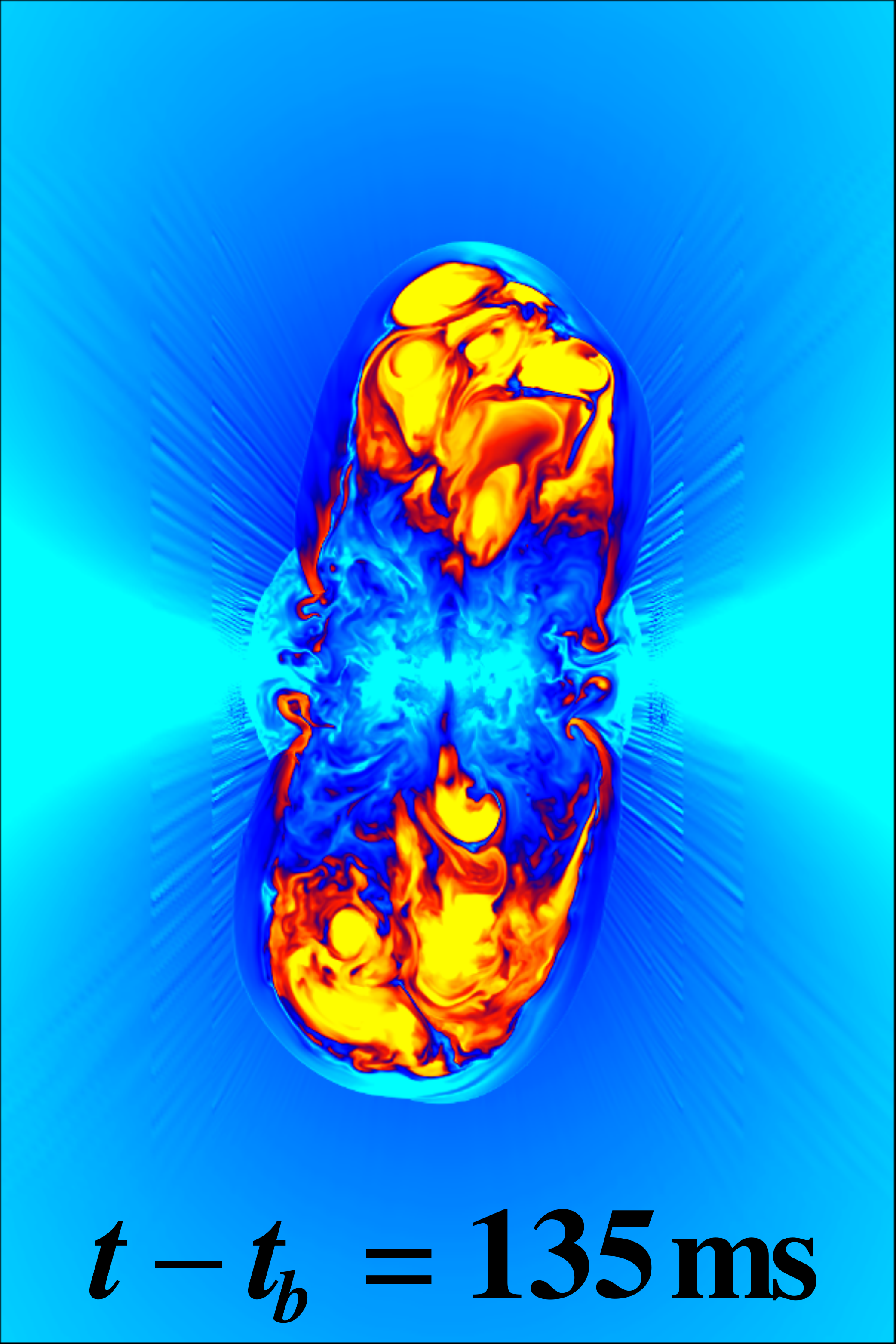}
\includegraphics[width=0.1625\textwidth]{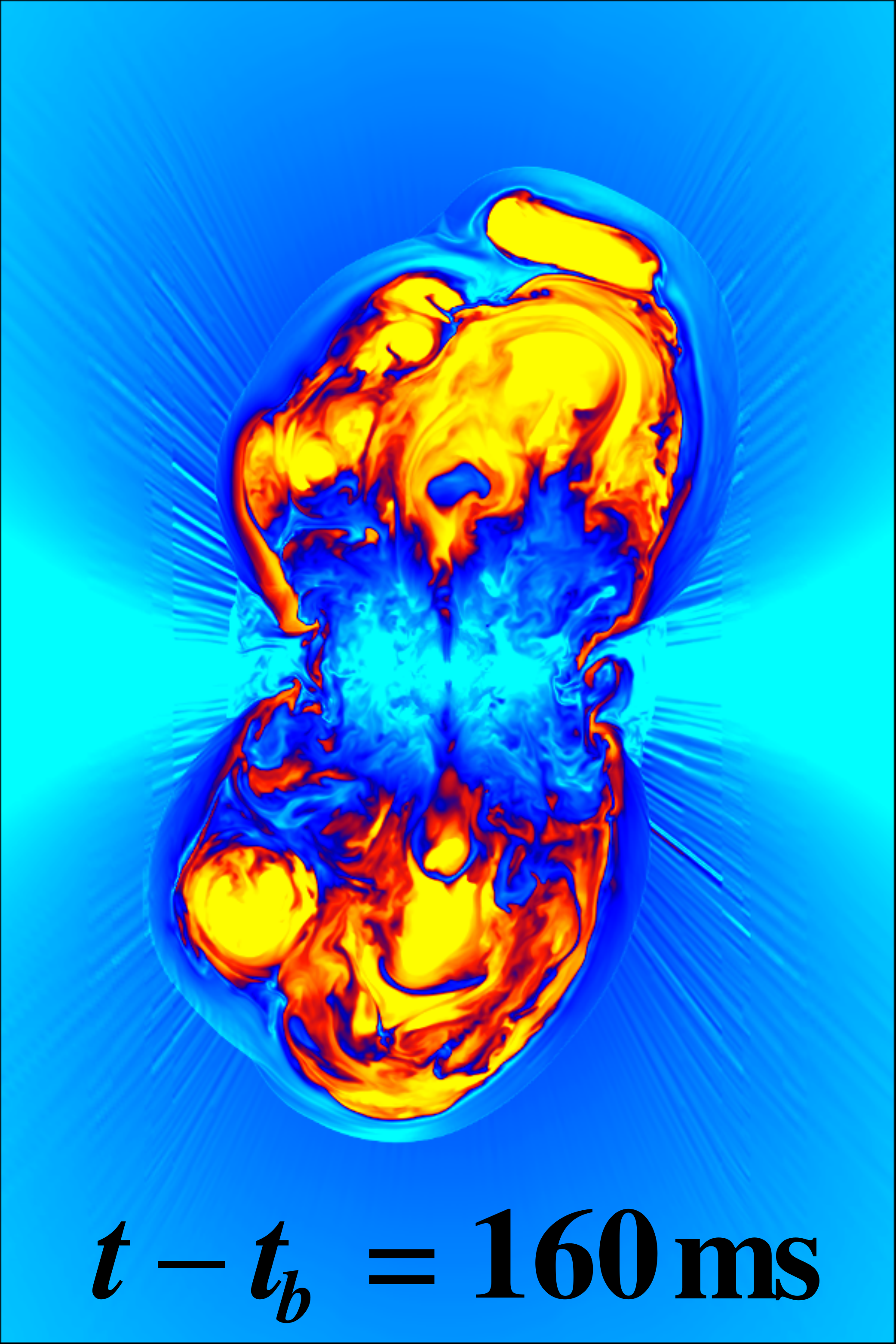}
\includegraphics[width=0.1625\textwidth]{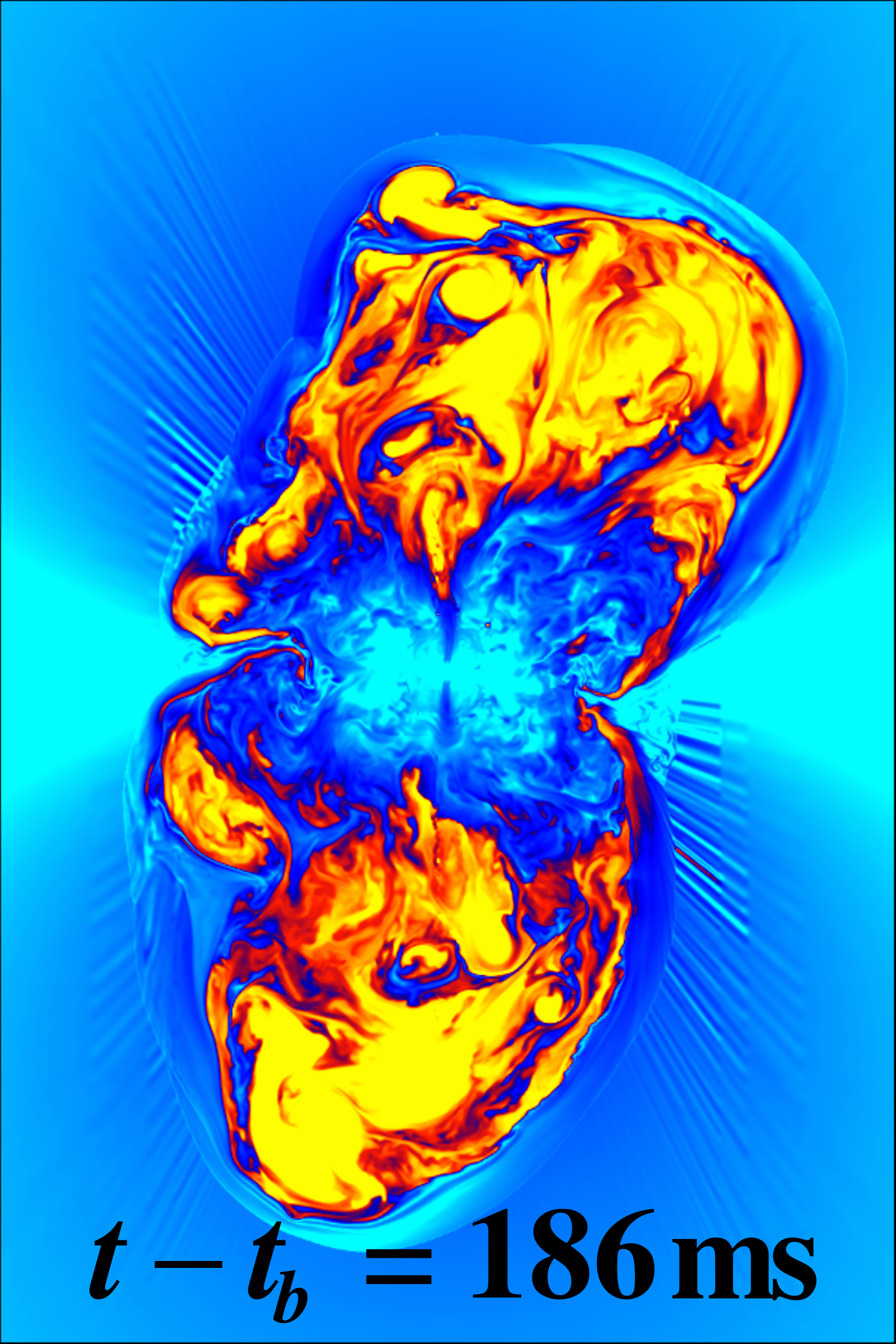}
\caption{Meridional slices ($x-z$ plane) of the plasma $\beta =
  P_\mathrm{gas}/P_\mathrm{mag}$ at different postbounce times. The
  $z$-axis of the frames is the vertical and the rotation axis, $x$ is
  the horizontal. The colormap is saturated at a minimum $\beta =
  0.01$ and a maximum $\beta = 10$.  Regions of $\beta < 1$ (warm
  colors, magnetically dominated) are underdense due to expansion from
  magnetic pressure, rise buoyantly, and push out the shock front
  in two prominent polar lobes.}
\label{fig:beta} \vspace{1.5ex} \vspace{0.5cm} \end{figure*}

\subsection{Nonaxisymmetric Instability and Jet Formation}
\label{sec:instabilities}

The results discussed in the above suggest that the full 3D run is
subject to a spiral instability that grows from $\sim$$1\%$ $m=1$
seeds in the velocity field to non-linear scale within the first
$\sim$$20\,\mathrm{ms}$ after bounce. This instability quenches the
jet. Figure~\ref{fig:growthrate} depicts the linear growth and
non-linear saturation of the spiral instability at various locations
along the spin axis outside the protoneutron star.

In the rotating CCSN context, rotational shear instabilities in the
protoneutron star (e.g., \citealt{ott:07prl}) and a spiral standing
accretion shock instability (SASI; e.g., \citealt{kuroda:13}) have been
discussed to potentially arise already at early postbounce
times.
It is unlikely that either of these is excited in our simulations, since
we choose to perturb a radially and vertically narrow region along the
spin axis outside of the protoneutron star and far from the shock,
 within the region of the highest toroidal magnetic field
strength. A spiral MHD instability may thus be the
driving agent behind the strong asymmetry in our simulation.

One possible such instability is the screw-pinch kink instability that has been
studied in jets from active galactic nuclei (e.g.,
\citealt{begelman:98,mignone:10,mckinney:09}).
The B-field near the spin axis in our simulation can be
roughly approximated by a screw-pinch field configuration. This
consists of a non-rotating plasma cylinder and a magnetic field of the
form 
\begin{equation} \vec{B} =
B_\mathrm{tor}(r)\hat{\phi} + B_z \hat{z}\,\,, 
\end{equation}
where $\hat{z}$ is along the rotation axis, $\hat{\phi}$ is in the
toroidal direction, $B_z$ is a constant vertical component of the
B-field, and $B_\mathrm{tor}(r)$ is the radially-varying toroidal
component of the B-field. We can express perturbations to the jet in
the form of fluid displacements as a sum of basis elements of the form
$\vec{\xi}_{km} \propto e^{i(kz + m\phi - \omega t)}$, where $m$ is an
integer, $k$ is the vertical wave number, and $\omega$ is the
oscillation frequency of the mode. The Kruskal-Shafranov stability
criterion states that a plasma cylinder confined to a finite radius
$a$ (as in a tokamak) is unstable to kink ($m=\pm1$) modes if
$B_\mathrm{tor}/B_z > 2 \pi a/L$, where L is the length of the
cylinder and the sign of $m$ is such that the mode's helicity is
opposite to the field helix
(\citealt{shafranov:56,kruskal:58}). Unconfined screw-pinch structures
with $B_\mathrm{tor} \gg B_z$ have been shown to be violently unstable
to $m=1$ modes at short vertical wavelengths ($kr \gg 1$) when $d\ln
B_\mathrm{tor}/d\ln r > -1/2$ and the plasma parameter is sufficiently large
($\beta > 2/3\gamma$ where $\gamma$ is the local adiabatic index). Under
these conditions (which are only approximately met in our simulation), the
fastest growing unstable mode (fgm) is amplified on a timescale comparable to the
Alfv\'en travel time around a toroidal loop \citep{begelman:98}. The expected
$m=1$ growth timescale and vertical wavelength in the most unstable regions of
the jet at $\sim$$10-15\, \mathrm{ms}$ after bounce are \begin{eqnarray*}
\tau_{\mathrm{fgm}} \approx \frac{4a\sqrt{\pi \rho}}{B_\mathrm{tor}} \approx
1\, \mathrm{ms}\,, \qquad \lambda_{\mathrm{fgm}} \approx \frac{4\pi
aB_z}{B_\mathrm{tor}} \approx 5\, \mathrm{km}\,, \end{eqnarray*} where $a$ is
the radius of the most unstable region.  The growth time is much shorter than
the time it would take for the jet to propagate through the shocked region.

The effect of the kink instability can be most clearly seen in a
displacement of the jet barycenter away from the rotation axis of
the core. We measure the displacement of the jet in our full 3D run by
computing the barycenter displacement (planar ``center-of-mass''
displacement; \citealt{mignone:10}) of the co-moving magnetic field
strength $b^2$ (see,
e.g., \citealt{moesta:14a}) in xy-slices at different heights $z_i$
along the rotation axis (Fig.~\ref{fig:growthrate}).
$b^2$ probes the MHD effects in our simulations most directly, but
other variable choices, e.g.\ the specific entropy $s$, exhibit
similar behavior as flux freezing couples fluid properties to the
magnetic field evolution.  Figure~\ref{fig:growthrate} demonstrates
that the jet experiences significant displacements from the rotation
axis of the core in a spiraling motion with helicity opposite to that
of the magnetic field (indicated by the magenta colored arrow in
Fig.~\ref{fig:growthrate}), and that the growth rate and dominant
instability length scale roughly agree with those predicted by a kink
unstable jet in our analysis.


\subsection{Magnetized Expanding Lobes}

Although the initial bipolar jet fails to promptly break out of the
stalled shock, MHD becomes dominant tens of milliseconds later.
Starting around $\sim$$80\, \mathrm{ms}$ after bounce, outflows of
highly-magnetized material are continuously launched from the
protoneutron star and propagate along the rotation axis of the core.
This is depicted in Fig.~\ref{fig:beta}, which presents meridional
slices of the plasma parameter $\beta$ at a range of postbounce times.
The highly-magnetized (low-$\beta$) high-entropy material does not
stay neatly confined to the rotation axis. 

\begin{figure*}[t] 
\centering
\includegraphics[width=0.49\textwidth]{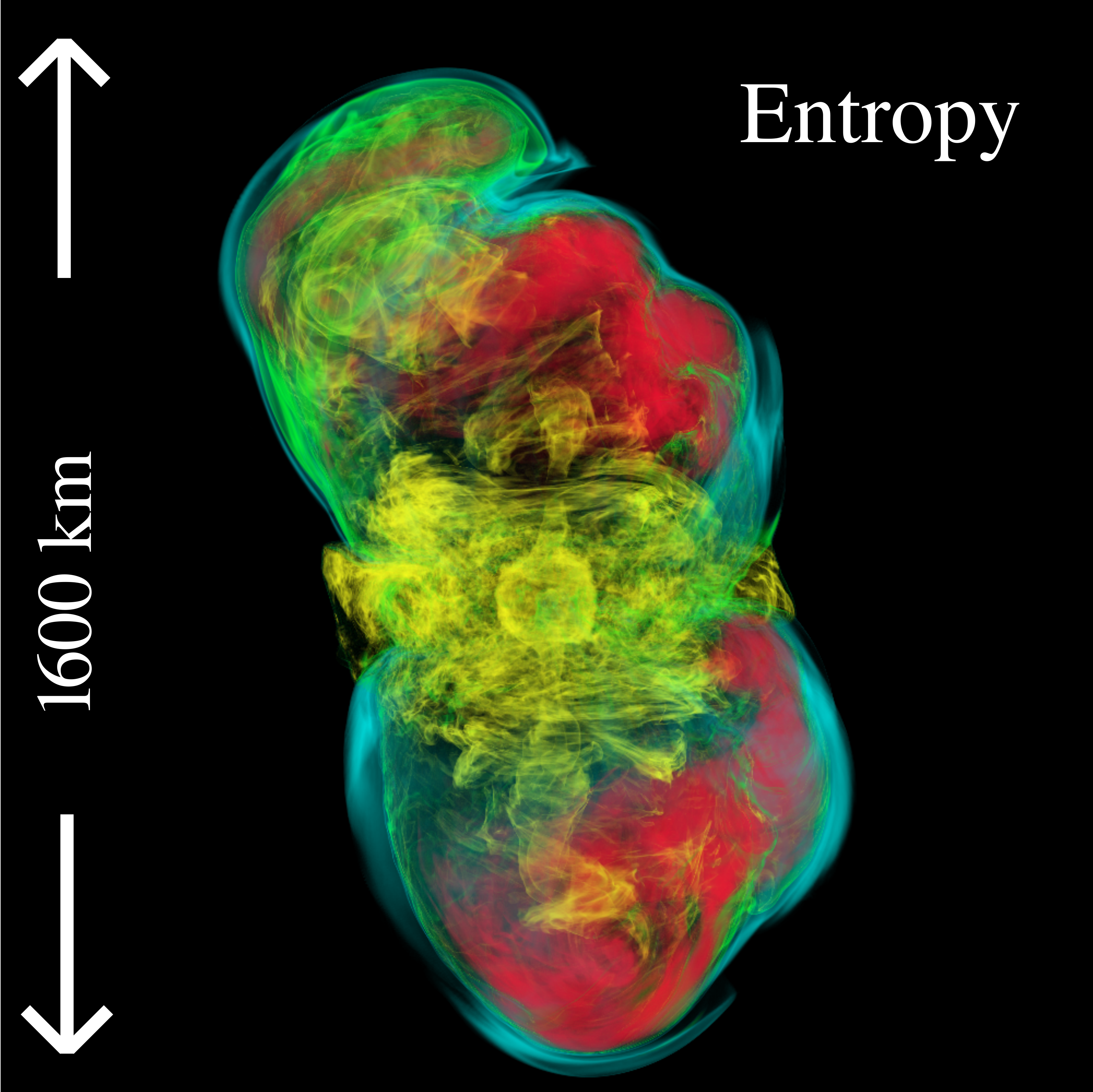}
\includegraphics[width=0.49\textwidth]{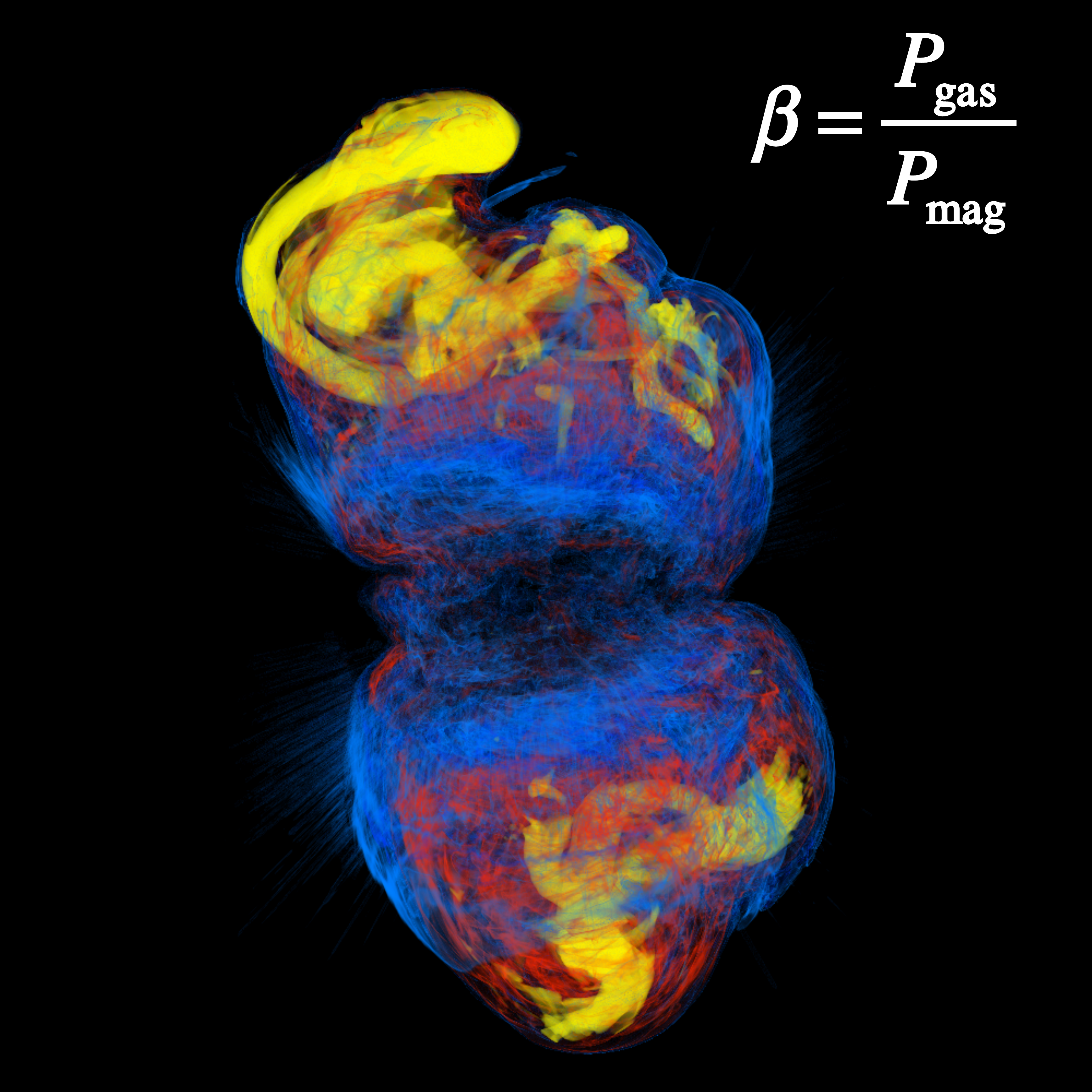}
\caption{Volume renderings of entropy and $\beta$ at $t-t_b = 161\,
  \mathrm{ms}$. The z-axis is the spin axis of the protoneutron
  star and we show $1600\,\mathrm{km}$ on a side.
  The colormap for entropy is chosen such that blue corresponds to $s
  = 3.7 k_b\, \mathrm{baryon}^{-1}$, cyan to $s = 4.8 k_b\,
  \mathrm{baryon}^{-1}$ indicating the shock surface, green to $s =
  5.8 k_b\, \mathrm{baryon}^{-1}$, yellow to $s = 7.4 k_b\,
  \mathrm{baryon}^{-1}$, and red to higher entropy material at $s = 10
  k_b\, \mathrm{baryon}^{-1}$. For $\beta$ we choose yellow to
  correspond to $\beta = 0.1$, red to $\beta = 0.6$, and blue to
  $\beta = 3.5$.  Magnetically dominated material at $\beta < 1$
  (yellow) is expelled from the protoneutron star and twisted in
  highly asymmetric tubes that drive the secular expansion of the
  polar lobes.} \label{fig:betavolren}
\vspace{1.5ex} \vspace{0.5cm} \end{figure*}

In Fig.~\ref{fig:betavolren}, we present volume renderings of specific
entropy and plasma parameter $\beta$ at $161\,\mathrm{ms}$ after
bounce. Only these volume renderings speak the full truth about how
severely outflows driven by the core are deformed, sheared, and wound
up as they propagate in the $z$-direction. The material that is
expelled from the vicinity of the protoneutron star forms tube-like
structures that are highly magnetized ($\beta \sim 0.01 - 0.1$),
underdense ($\sim$$1\%$ the density of the surrounding fluid), and rise
buoyantly. The overall structure of the shocked region evolves toward strongly
prolate-shape with two, roughly $90^\circ$-filling tilted lobes at both poles
(cf.~Figs.~\ref{fig:octfullcmp}, \ref{fig:beta}, \ref{fig:betavolren}). 

The lobes secularly expand to $\sim$$900\,\mathrm{km}$ during the
simulated time, but their expansion never becomes dynamical. Accreting
material is funneled to equatorial regions where it continues to
settle onto the protoneutron star. The $B$-field geometry in the later
evolution corresponds to that of a tightly wound coil close to the
protoneutron star, but the field lines open up in a spiraling fashion
further out, yet still behind the shock. This is consistent with
magnetized material moving away from the rotation axis as it
propagates in the general $z$-direction. In 2D simulations, a
confining magnetic-tower structure forms instead \citep{burrows:07b}.

\section{Discussion}



Our results show that 3D magnetorotational core-collapse supernovae
are fundamentally different from what has been anticipated on the
basis of axisymmetric simulations
\citep{burrows:07b,dessart:08a,takiwaki:11}.  A jet that develops in
2D is disrupted and fizzles in 3D.  We suggest that the instability
driving this is most likely an MHD kink ($m=1$) instability to which
the toroidally-dominated postbounce magnetic-field configuration is
prone. Instead of an axially symmetric jet, a completely new,
previously unseen, wide double-lobed flow pattern develops, but
we obtain no runaway explosion during the simulated time.   


The high precollapse field strength of $10^{12}\,\mathrm{G}$ yields
$\sim$$10^{16}\,\mathrm{G}$ in toroidal field and $\beta =
P_\mathrm{gas}/P_\mathrm{mag} < 1$ within only
$\sim$$10-15\,\mathrm{ms}$ of bounce, creating conditions favorable
for jet formation. Yet, the growth time of the kink instability
is shorter than the time it takes for the jet to develop.  In a
short test simulation with an even more unrealistic, ten times
stronger initial field, a successful jet is launched promptly after
bounce (consistent with \citealt{winteler:12}), but
subsequently experiences a spiral instability in its core. 

Realistic precollapse iron cores are not expected to have magnetic fields
in excess of $\sim$$10^{8}-10^{9}\,\mathrm{G}$, which may be amplified to no
more than $\sim$$10^{12}\,\mathrm{G}$ during collapse
\citep{burrows:07b}. The $10^{15}-10^{16}\,\mathrm{G}$ of large-scale
toroidal field required to drive a magnetorotational jet must be built
up after bounce. This will likely require tens to hundreds of
dynamical times, even if the magnetorotational instability operates in
conjunction with a dynamo. The results of the
present and previous full 3D rotating CCSN simulations
\citep{ott:07prl,kuroda:13} suggest that MHD and also a variety of
nonaxisymmetric hydrodynamic instabilities will grow to non-linear regimes on
\emph{shorter timescales}, disrupting any possibly developing axial outflow.
\emph{This is why we believe
  that the dynamics and flow structures seen in our full 3D simulation
  may be generic to the postbounce evolution of rapidly rotating
  magnetized core collapse that starts from realistic initial
  conditions}. 

If the polar lobes eventually accelerate, the resulting explosion will
be asymmetric, though probably less so than a jet-driven
explosion. The lobes carry neutron rich ($Y_e \sim 0.1-0.2$) material
of moderate entropy ($s \sim 10-15\,k_B\,\mathrm{baryon}^{-1}$), which
could lead to interesting $r$-process yields, similar to what
\cite{winteler:12} found for their prompt jet-driven explosion.  Even
if the lobes continue to move outward, accretion in equatorial regions
may continue, eventually (after $2-3\,\mathrm{s}$) leading to the
collapse of the protoneutron star and black hole formation.  In this
case, the engine supplying the lobes with low-$\beta$ plasma is shut
off. Unless their material has reached positive total energy, the
lobes will fall back onto the black hole, which will subsequently
hyperaccrete until material becomes centrifugally supported in an
accretion disk. This would set the stage for a subsequent long GRB and an
associated Type Ic-bl CCSN that would be driven by a collapsar central
engine \citep{woosley:93} rather than by a protomagnetar \citep{metzger:11}.

\vskip.2cm The results of the present study highlight the importance
of studying magnetorotational CCSNe in 3D. Future work will be
necessary to explore later postbounce dynamics, the sensitivity to
initial conditions and numerical resolution, and possible
nucleosynthetic yields. Animations and further details on
our simulations are available at \url{http://stellarcollapse.org/cc3dgrmhd}.

\section*{Acknowledgments}

The authors would like to thank A.~Burrows, S.~Couch, U.~Gamma,
D.~Meier, and L.~Roberts for discussions. This research was partially
supported by NSF grants AST-1212170, PHY-1151197, and OCI-0905046, and
the Sherman Fairchild Foundation. SR is supported by a DOE
Computational Science Graduate Fellowship DE-FG02-97ER25308. CR
acknowledges support by NASA through Einstein Fellowship grant 
PF2-130099. The simulations were carried out on XSEDE (TG-PHY100033)
and on NSF/NCSA BlueWaters (PRAC OCI-0941653).

\begin{thebibliography}{43}
\expandafter\ifx\csname natexlab\endcsname\relax\def\natexlab#1{#1}\fi

\bibitem[{{Akiyama} {et~al.}(2003){Akiyama}, {Wheeler}, {Meier}, \&
  {Lichtenstadt}}]{akiyama:03}
{Akiyama}, S., {Wheeler}, J.~C., {Meier}, D.~L., \& {Lichtenstadt}, I. 2003,
  \apj, 584, 954

\bibitem[{{Balbus} \& {Hawley}(1991)}]{balbus:91}
{Balbus}, S.~A., \& {Hawley}, J.~F. 1991, \apj, 376, 214

\bibitem[{{Begelman}(1998)}]{begelman:98}
{Begelman}, M.~C. 1998, \apj, 493, 291

\bibitem[{{Bethe} \& {Wilson}(1985)}]{bethe:85}
{Bethe}, H.~A., \& {Wilson}, J.~R. 1985, \apj, 295, 14

\bibitem[{{Bisnovatyi-Kogan}(1970)}]{bisno:70}
{Bisnovatyi-Kogan}, G.~S. 1970, Astron. Zh., 47, 813

\bibitem[{{Bruenn} {et~al.}(2013){Bruenn}, {Mezzacappa}, {Hix}, {Lentz},
  {Bronson Messer}, {Lingerfelt}, {Blondin}, {Endeve}, {Marronetti}, \&
  {Yakunin}}]{bruenn:13}
{Bruenn}, S.~W., {Mezzacappa}, A., {Hix}, W.~R., {et~al.} 2013, \apjl, 767, L6

\bibitem[{{Burrows} {et~al.}(2007){Burrows}, {Dessart}, {Livne}, {Ott}, \&
  {Murphy}}]{burrows:07b}
{Burrows}, A., {Dessart}, L., {Livne}, E., {Ott}, C.~D., \& {Murphy}, J. 2007,
  \apj, 664, 416

\bibitem[{{Dessart} {et~al.}(2008){Dessart}, {Burrows}, {Livne}, \&
  {Ott}}]{dessart:08a}
{Dessart}, L., {Burrows}, A., {Livne}, E., \& {Ott}, C.~D. 2008, \apjl, 673,
  L43

\bibitem[{{Drout} {et~al.}(2011){Drout}, {Soderberg}, {Gal-Yam}, {Cenko},
  {Fox}, {Leonard}, {Sand}, {Moon}, {Arcavi}, \& {Green}}]{drout:11}
{Drout}, M.~R., {Soderberg}, A.~M., {Gal-Yam}, A., {et~al.} 2011, \apj, 741, 97

\bibitem[{{Einfeldt}(1988)}]{HLLE:88}
{Einfeldt}, B. 1988, in Shock tubes and waves; Proceedings of the Sixteenth
  International Symposium, Aachen, Germany, July 26--31, 1987. VCH Verlag,
  Weinheim, Germany, 671

\bibitem[{{Heger} {et~al.}(2000){Heger}, {Langer}, \& {Woosley}}]{heger:00}
{Heger}, A., {Langer}, N., \& {Woosley}, S.~E. 2000, \apj, 528, 368

\bibitem[{{Hjorth} \& {Bloom}(2011)}]{hjorth:11}
{Hjorth}, J., \& {Bloom}, J.~S. 2011, in Gamma-Ray Bursts, ed. C.~Kouveliotou,
  R.~A. M.~J. Wijers, \& S.~E. Woosley (Cambridge, UK: Cambridge University
  Press)

\bibitem[{{Kruskal} \& {Tuck}(1958)}]{kruskal:58}
{Kruskal}, M., \& {Tuck}, J.~L. 1958, Proc. R. Soc. Lond., 245, 222

\bibitem[{{Kuroda} {et~al.}(2013){Kuroda}, {Takiwaki}, \& {Kotake}}]{kuroda:13}
{Kuroda}, T., {Takiwaki}, T., \& {Kotake}, K. 2013, Submitted to Phys.~Rev.~D.,
  arXiv:1304.4372

\bibitem[{{Kuroda} \& {Umeda}(2010)}]{kuroda:10}
{Kuroda}, T., \& {Umeda}, H. 2010, \apjs, 191, 439

\bibitem[{Lattimer \& Swesty(1991)}]{lseos:91}
Lattimer, J.~M., \& Swesty, F.~D. 1991, {Nucl. Phys. A}, 535, 331

\bibitem[{{LeBlanc} \& {Wilson}(1970)}]{leblanc:70}
{LeBlanc}, J.~M., \& {Wilson}, J.~R. 1970, \apj, 161, 541

\bibitem[{{L{\"o}ffler} {et~al.}(2012){L{\"o}ffler}, {Faber}, {Bentivegna},
  {Bode}, {Diener}, {Haas}, {Hinder}, {Mundim}, {Ott}, {Schnetter}, {Allen},
  {Campanelli}, \& {Laguna}}]{loeffler:12}
{L{\"o}ffler}, F., {Faber}, J., {Bentivegna}, E., {et~al.} 2012, Class. Quantum
  Grav., 29, 115001

\bibitem[{{McKinney} \& {Blandford}(2009)}]{mckinney:09}
{McKinney}, J.~C., \& {Blandford}, R.~D. 2009, \mnras, 394, L126

\bibitem[{{Meier} {et~al.}(1976){Meier}, {Epstein}, {Arnett}, \&
  {Schramm}}]{meier:76}
{Meier}, D.~L., {Epstein}, R.~I., {Arnett}, W.~D., \& {Schramm}, D.~N. 1976,
  \apj, 204, 869

\bibitem[{{Metzger} {et~al.}(2011){Metzger}, {Giannios}, {Thompson},
  {Bucciantini}, \& {Quataert}}]{metzger:11}
{Metzger}, B.~D., {Giannios}, D., {Thompson}, T.~A., {Bucciantini}, N., \&
  {Quataert}, E. 2011, \mnras, 413, 2031

\bibitem[{{Mignone} {et~al.}(2010){Mignone}, {Rossi}, {Bodo}, {Ferrari}, \&
  {Massaglia}}]{mignone:10}
{Mignone}, A., {Rossi}, P., {Bodo}, G., {Ferrari}, A., \& {Massaglia}, S. 2010,
  \mnras, 402, 7

\bibitem[{{Mikami} {et~al.}(2008){Mikami}, {Sato}, {Matsumoto}, \&
  {Hanawa}}]{mikami:08}
{Mikami}, H., {Sato}, Y., {Matsumoto}, T., \& {Hanawa}, T. 2008, \apj, 683, 357

\bibitem[{{Modjaz}(2011)}]{modjaz:11}
{Modjaz}, M. 2011, Astron. Nachr., 332, 434

\bibitem[{{M{\"o}sta} {et~al.}(2014){M{\"o}sta}, {Mundim}, {Faber}, {Haas},
  {Noble}, {Bode}, {L{\"o}ffler}, {Ott}, {Reisswig}, \&
  {Schnetter}}]{moesta:14a}
{M{\"o}sta}, P., {Mundim}, B.~C., {Faber}, J.~A., {et~al.} 2014, Class. Quantum
  Grav., 31, 015005

\bibitem[{{M{\"u}ller} {et~al.}(2012){M{\"u}ller}, {Janka}, \&
  {Marek}}]{mueller:12a}
{M{\"u}ller}, B., {Janka}, H.-T., \& {Marek}, A. 2012, \apj, 756, 84

\bibitem[{{Obergaulinger} {et~al.}(2009){Obergaulinger}, {Cerd{\'a}-Dur{\'a}n},
  {M{\"u}ller}, \& {Aloy}}]{obergaulinger:09}
{Obergaulinger}, M., {Cerd{\'a}-Dur{\'a}n}, P., {M{\"u}ller}, E., \& {Aloy},
  M.~A. 2009, \aap, 498, 241

\bibitem[{{O'Connor} \& {Ott}(2010)}]{oconnor:10}
{O'Connor}, E., \& {Ott}, C.~D. 2010, Class. Quantum Grav., 27, 114103

\bibitem[{{Ott} {et~al.}(2006){Ott}, {Burrows}, {Thompson}, {Livne}, \&
  {Walder}}]{ott:06spin}
{Ott}, C.~D., {Burrows}, A., {Thompson}, T.~A., {Livne}, E., \& {Walder}, R.
  2006, Astrophys. J. Suppl. Ser., 164, 130

\bibitem[{{Ott} {et~al.}(2007){Ott}, {Dimmelmeier}, {Marek}, {Janka}, {Hawke},
  {Zink}, \& {Schnetter}}]{ott:07prl}
{Ott}, C.~D., {Dimmelmeier}, H., {Marek}, A., {et~al.} 2007, \prl, 98, 261101

\bibitem[{{Ott} {et~al.}(2012){Ott}, {Abdikamalov}, {O'Connor}, {Reisswig},
  {Haas}, {Kalmus}, {Drasco}, {Burrows}, \& {Schnetter}}]{ott:12a}
{Ott}, C.~D., {Abdikamalov}, E., {O'Connor}, E., {et~al.} 2012, \prd, 86,
  024026

\bibitem[{{Reisswig} {et~al.}(2013){Reisswig}, {Haas}, {Ott}, {Abdikamalov},
  {M{\"o}sta}, {Pollney}, \& {Schnetter}}]{reisswig:12}
{Reisswig}, C., {Haas}, R., {Ott}, C.~D., {et~al.} 2013, \prd, 87, 064023

\bibitem[{{Scheidegger} {et~al.}(2010){Scheidegger}, {K{\"a}ppeli},
  {Whitehouse}, {Fischer}, \& {Liebend{\"o}rfer}}]{scheidegger:10b}
{Scheidegger}, S., {K{\"a}ppeli}, R., {Whitehouse}, S.~C., {Fischer}, T., \&
  {Liebend{\"o}rfer}, M. 2010, \aap, 514, A51

\bibitem[{{Shafranov}(1956)}]{shafranov:56}
{Shafranov}, V. 1956, At. Energy, 5, 38

\bibitem[{{Shibata} {et~al.}(2006){Shibata}, {Liu}, {Shapiro}, \&
  {Stephens}}]{shibata:06}
{Shibata}, M., {Liu}, Y.~T., {Shapiro}, S.~L., \& {Stephens}, B.~C. 2006, \prd,
  74, 104026

\bibitem[{{Soderberg} {et~al.}(2006){Soderberg}, {Kulkarni}, {Nakar}, {Berger},
  {Cameron}, {Fox}, {Frail}, {Gal-Yam}, {Sari}, {Cenko}, {Kasliwal},
  {Chevalier}, {Piran}, {Price}, {Schmidt}, {Pooley}, {Moon}, {Penprase},
  {Ofek}, {Rau}, {Gehrels}, {Nousek}, {Burrows}, {Persson}, \&
  {McCarthy}}]{soderberg:06}
{Soderberg}, A.~M., {Kulkarni}, S.~R., {Nakar}, E., {et~al.} 2006, \nat, 442,
  1014

\bibitem[{{Takiwaki} \& {Kotake}(2011)}]{takiwaki:11}
{Takiwaki}, T., \& {Kotake}, K. 2011, \apj, 743, 30

\bibitem[{Tchekhovskoy {et~al.}(2007)Tchekhovskoy, McKinney, \&
  Narayan}]{tchekhovskoy:07}
Tchekhovskoy, A., McKinney, J.~C., \& Narayan, R. 2007, Mon. Not. Roy. Astron.
  Soc., 379, 469

\bibitem[{{Thompson} {et~al.}(2005){Thompson}, {Quataert}, \&
  {Burrows}}]{thompson:05}
{Thompson}, T.~A., {Quataert}, E., \& {Burrows}, A. 2005, \apj, 620, 861

\bibitem[{{T{\'o}th}(2000)}]{toth:00}
{T{\'o}th}, G. 2000, J. Comp. Phys., 161, 605

\bibitem[{{Wheeler} {et~al.}(2002){Wheeler}, {Meier}, \& {Wilson}}]{wheeler:02}
{Wheeler}, J.~C., {Meier}, D.~L., \& {Wilson}, J.~R. 2002, \apj, 568, 807

\bibitem[{{Winteler} {et~al.}(2012){Winteler}, {K{\"a}ppeli}, {Perego},
  {Arcones}, {Vasset}, {Nishimura}, {Liebend{\"o}rfer}, \&
  {Thielemann}}]{winteler:12}
{Winteler}, C., {K{\"a}ppeli}, R., {Perego}, A., {et~al.} 2012, \apjl, 750, L22

\bibitem[{{Woosley}(1993)}]{woosley:93}
{Woosley}, S.~E. 1993, \apj, 405, 273

\end{thebibliography}

\end{document}